\documentclass[prl,twocolumn,epsf]{revtex4-1}
\usepackage{graphicx}
\usepackage{dcolumn}
\usepackage{bm}
\usepackage{hyperref}

\newcommand{\beq}{\begin{eqnarray}}
\newcommand{\eeq}{\end{eqnarray}}
\begin{document}
\noindent
{\bf Comment on ``Measurement of x-ray absorption spectra of overdoped high-temperature cuprate superconductors: Inapplicability of the single-band Hubbard model '' }

\begin{figure}
\centering
\includegraphics[width=9.0cm]{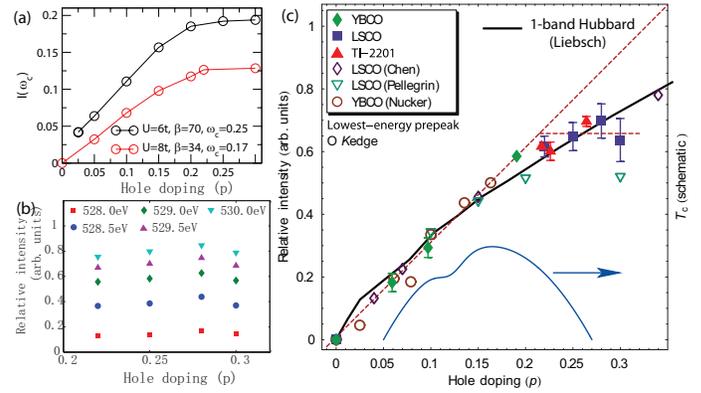}
\caption{a) Integrated densities of states on the interval $[0,\omega_c]$, for $U=8t$ and $U=6t$. b) Integrated oxygen K-edge intensity 
of the data of Peets, et al.\cite{peets} for LSCO with the cutoffs shown
explicitly.  c) Fig. (2) of Peets, et al.\cite{peets} overlayed (solid line) with the simulation data (Fig. 2a of Liebsch\cite{liebsch}) on the 1-band Hubbard model with an appropriately chosen y-axis
scale factor to account for the mismatch with the units used in the experimental data. The dashed lines are due to Peets, et al.\cite{peets}.}
\label{fig1}
\end{figure}

In a recent Letter,  Peets, et al.\cite{peets} measured the x-ray
intensity at the oxygen K-edge (hereafter referred to as LESW) in overdoped
La$_{2-x}$Sr$_x$CuO$_{4\pm\delta}$ (LSCO) and
Tl$_2$Ba$_2$CuO$_{6+\delta}$.  They concluded that, unlike the
underdoped samples of LSCO and YBa$_2$Cu$_3$O$_x$  in which LESW increases at least linearly\cite{chen} with
doping, it saturates (see dashed lines in Fig. (\ref{fig1})c) abruptly for a hole count exceeding $x_c\approx
0.23$.   They interpreted\cite{peets} the saturation as a breakdown of
the 1-band Hubbard model in the cuprates. We analyse their data and
show that this conclusion does not follow necessarily.

To clarify, the 1-band Hubbard model is designed to capture the low-energy
features of the cuprates, particularly on the pseudogap scale.  In
fact, as demonstrated previously\cite{3band}, beyond an energy scale of 500meV,
deviations with the physics of the 3-band model are
noticeable\cite{3band}. The oxygen K-edge experiments measure the
unoccupied part of the density of states projected onto the oxygen 2p
states.  Hence, the relevant question is, can the 1-band
Hubbard model reproduce the purported saturation with a cutoff on the
integrated density of states of no more than 500meV?  Indeed it can as
shown in Fig. (\ref{fig1}a).  Shown here is a calculation of the
integrated density of states using  the
dynamical cluster approximation (DCA), with a quantum Monte Carlo
algorithm as the cluster solver\cite{hettler}.  We used a 16-site
cluster. The densities of states produced by this method all show
pseudogaps for the low-doping regime. Focusing entirely on the
low-energy PG physics, we integrated the density of states up to a
cutoff of $2J$.  The calculations are for the lowest accessible temperatures, but the main feature that the integrated intensity levels once the pseudogap closes ($x=0.13$ for $U=6t$ and $x=0.22$  for $U=8t$) persists up to $T\sim J$ as is seen experimentally\cite{peets}. This is not an accident.  When the pseudogap
closes, Fermi liquid behaviour ensues, making it meaningless to
separate the spectrum above the chemical potential into high and low energy parts.   

Fig. (\ref{fig1}a) is entirely {\bf illustrative} because of course,
{\bf no} saturation obtains in the 1-band Hubbard model\cite{eskes,phillips,liebsch}
if the cutoff
encompasses the
 full LESW.  But is the choice of the cutoff (and hence the purported deviation
from the full weight in the 1-band model) relevant to the experiments? Fig. (\ref{fig1}b) shows that it is not because regardless of the cutoff used to determine the LESW for LSCO, the saturation persists. Hence, the purported saturation in the
experiments, if it truly exists, is unrelated to any high-energy scale. In fact, the width of integration with a cutoff of $528.5eV$ is identical to the integration range in the artificial calculation in Fig. (\ref{fig1}a). 

Nonetheless, it is difficult to reconcile a saturation of the full LESW
with any known
model.  To investigate how robust the claim of saturation is, we overlay the recent cluster data (solid line in Fig. (\ref{fig1}c)) of Liebsch\cite{liebsch} of the LESW evaluated
with a doping-dependent cutoff that excludes any contribution from the upper band.  The agreement with the data points (error bars included) is excellent.  The slope change in the LESW in the 1-band Hubbard model reflects the fact that the dynamical spectral weight transfer\cite{eskes,phillips} (the explicit $t/U$ corrections which make the LESW per spin exceed the doping level, $x$)
 must diminish above a certain doping level so that the non-interacting value of one state per
site per spin is recovered at $n=0$.  Consequently, the claim that the
experimental data (with error bars)  imply a failure of the 1-band Hubbard model is not substantiated. 

\vspace{12pt}
\noindent Philip Phillips,
Loomis Laboratory of Physics\\
University of Illinois,
Urbana, Il. 61801-3080\\
Mark Jarrell,
Department of Physics and Astronomy,\\
Louisianna State University,
Baton Rouge, LA., 70803

\acknowledgements We acknowledge the NSF DMR-0940992 and DMR-0706379 for partial support.   This research used NCCS resources at ORNL, supported by DOE Contract No. DE-AC05-00OR22725.

\end{document}